\documentstyle[epsf,prb,multicol,aps]{revtex}
\begin{document}

\draft

\title{Electron Dynamics in Intentionally Disordered Semiconductor
Superlattices}

\author{Enrique Diez and Angel S\'anchez }

\address{Departamento de Matem\'aticas,
Escuela Polit\'ecnica Superior,
Universidad Carlos III, %de Madrid,
E-28911 Legan\'es, Madrid, Spain}

\author{Francisco Dom\'{\i}nguez-Adame}

\address{Departamento de F\'{\i}sica de Materiales,
Facultad de F\'{\i}sicas, Universidad Complutense,
E-28040 Madrid, Spain}

\author{Gennady P.\ Berman}

\address{Theoretical Division and CNLS, B213, Los Alamos National
Laboratory, Los Alamos, New Mexico 87545}

\maketitle

\begin{abstract}

We study the dynamical behavior of disordered quantum-well-based
semiconductor su\-per\-lattices where the disorder is intentional and
short-range correlated.  We show that, whereas the transmission time of
a particle grows exponentially with the number of wells in an usual
disordered superlattice for any value of the incident particle energy,
for specific values of the incident energy this time increases linearly
when correlated disorder is included.  As expected, those values of the
energy coincide with a narrow subband of extended states predicted by
the static calculations of Dom\'{\i}nguez-Adame {\em et al.}\ [Phys.\
Rev.\ B {\bf 51}, 14\,359 (1994)]; such states are seen in our dynamical
results to exhibit a ballistic regime, very close to the WKB
approximation of a perfect super\-la\-ttice.  Fourier transform of the
output signal for an incident Gaussian wave packet reveals a dramatic
filtering of the original signal, which makes us confident that devices
based on this property may be designed and used for nanotechnological
applications.  This is more so in view of the possibility of controlling
the output band using a dc electric field, which we also discuss.  In
the conclusion we summarize our results and present an outlook for
future developments arising from this work.

\end{abstract}

    %  Nanotechnology

\makeatletter
\global\@specialpagefalse
\def\@oddhead{\underline{\em To be published in Physical Review B \hspace{271pt}
$\sim$November 1996.}}
\let\@evenhead\@oddhead
\makeatother

\begin{multicols}{2}

\narrowtext

\section{Introduction}
\label{sec1}

In recent years, there has been a growing interest in studies of
disordered systems where the disorder presents some kind of correlation
(see Ref.\ \onlinecite{kike1} and references therein).  Aiming to find a
physically realizable system of this type, S\'anchez and
Dom\'{\i}nguez-Adame developed a simplified, continuous model in Ref.\
\onlinecite{anxo1} for studying disordered semiconductor superlattices
(SL's) where the disorder exhibits short-range spatial correlations.  In
this particular class of disordered SL's bands of extended states
appear, opposite to the conventional view that in one-dimensional (1D)
random systems almost all eigenstates are exponentially localized (see,
e.g., Ref.\ \onlinecite{Ziman}).  Much more realistic calculations
proved that these extended states are relevant to transport properties
of actual superlattices, giving rise to large DC conductivities when the
Fermi energy lies in one of these bands.\cite{ieee} However, all those
studies were carried out from a purely static viewpoint, and provided no
information about the dynamics of electrons in this new type of
nanostructures.

In view of the lack of this kind of analysis, we undertook the study of
the dynamical properties of electrons in these systems to complete the
static picture, already quite thorough.  Thus, we compute the behavior
of a wave packet incident on an intentionally disordered semiconductor
SL by numerically solving the 1D time-dependent Schr\"odinger equation
for the complete Hamiltonian (i.e., without analytical approximations)
in the presence of an electric field.  We explore several dynamical
characteristics of our system, such as the tunneling times and the
relation between the dwell time and the density of
states.\cite{buti1,Gennady,Iannacone} In addition, we estimate the
characteristic time over which the resonant quasi-level can be
established, showing that it is sufficiently large to allow the wave
packet to tunnel close to the ballistic regime.  We also consider the
competition between quantum coherence, preserved by correlated disorder,
and the loss of quantum coherence due to an electric field acting on the
SL. It is important to clarify that loss of quantum coherence\cite{Mendez}
means in this context any elastic processes causing a
complete localization of electronic states since we are not considering
dissipative processes.
Finally, we study the filter-like properties of these systems using
the Fourier transform of the {\em transmitted} part of the wave packet
and its dependence of the electric field, obtaining that it is possible
to control the width and the center of the filtered band.  It goes
without saying that a correct understanding of these properties is
crucial from the perspective of technological applications of
intentionally disordered SL's.

The paper is organized as follows.  In Sec.\ \ref{sec2} we present our
model and summarize previous work of us,\cite{ieee,quasi} which we find
convenient for a better understanding of the present paper, specifically
as regards the behavior of the transmission coefficient, with and
without electric field, for correlated and uncorrelated disordered SL's.
The body of the paper is Sec.\ \ref{sec4} where we present our dynamical
study of the system.  We begin by examining the transmission probability
and the transmission time for the two different kinds of SL's.  We
compute the dependence of the transmission time with the size of the
system in the WKB approximation for the ballistic regime and compare it
with the numerical results.  Most of the section is devoted to the
relation between the mean dwell time and the density of states and, in
addition, to the physical significance of the dwell time in this class
of disordered systems.  We complete this characterization with a study
of the spreading of the wave packet as a function of time.  Following
this equilibrium analysis, we devote Sec.\ \ref{sec5} to the study of
the effects produced by the electric field on the quantities presented
in the last section, placing particular emphasis on the the filtering
properties of the correlated disordered SL's.  Finally, in Sec.\
\ref{sec6}, we discuss our results and how can these be related to
actual measurements to infer the main characteristics of the bands of
the theoretically predicted extended states from experiments on SL's.
We close the paper with a few prospects on future developments that may
be attained starting from the present results.

\section{Model and Background}
\label{sec2}

\begin{figure}
\setlength{\epsfxsize}{7.0cm}
\centerline{\mbox{\epsffile{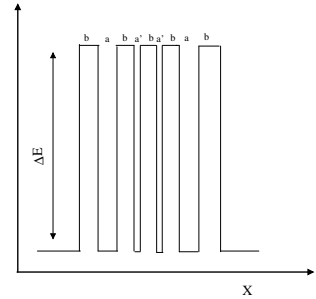}}}
\caption{Schematic diagram of the conduction-band profile of a SL
containing a DQW.}
\label{fig1}
\end{figure}

We resume in this section previous results of us\cite{ieee,quasi} for
correlated disordered SL's in the stationary case, which will be useful
for the discussion of the dynamical properties which we address in the
next section.  For our present purposes, it is enough to focus on
electron states close to the band gap with ${\bf k}_{||}=0$ and use the
one-band effective-mass framework to calculate the envelope-functions
\begin{equation}
\label{Schr}
\left[-\,{\hbar^2\over 2m^*}\,{d^2\phantom{x}\over dx^2} +
V_{\text{SL}}(x) - eF \, x \right] \psi(x) = E\>\psi(x),
\end{equation}
where a explicit dependence of both $E$ and $\psi(x)$ on quantum numbers
is understood and they will be omitted in the rest of the paper.  We
have taken a constant effective-mass $m^*$ at the $\Gamma$ valley
although this is not a serious limitation as our description can be
easily generalized to include two different effective masses.  In the
simplest picture, the SL potential $V_{\text{SL}}$ derives directly from
the different energies of the conduction- and valence-band edges at the
interfaces.  A single quantum-well (QW) consists of a layer of thickness
$d_A$ of a semiconductor A embedded in a semiconductor B. In our model
of disordered SL, we consider that $d_A$ takes at random one of two
values, $a$ and $a^{\prime}$.  We call this a random SL (RSL).  The
thickness of layers B separating neighboring QWs is assumed to be the
same in the whole SL, $d_B=b$.  A random dimer SL (DSL) is
built\cite{ieee} by imposing the additional constraint that QWs of
thickness $a^{\prime}$ appear only in pairs, called hereafter a dimer QW
(DQW), as shown in Fig.\ \ref{fig1}.  As a typical SL we have chosen a
GaAs-Ga$_{0.65}$Al$_{0.35}$As structure.  In this case, the
conduction-band offset is $\Delta E_c=0.25\,$eV, and the effective mass
is $m^*=0.067\,m$, $m$ being the electron mass. The origin of energies
is taken at the GaAs conduction-band edge. In our computations we
have taken $a=b=32\,$\AA\ and $a^{\prime}=26\,$\AA.  The fraction of QWs
of thickness $a^{\prime}$ is $40\,\%$ of the total number of QWs of the
SL. This is not an essential parameter of the model as similar results
are obtained taking other fractions.

\begin{figure}
\setlength{\epsfxsize}{4.0cm}
\centerline{\mbox{\epsffile{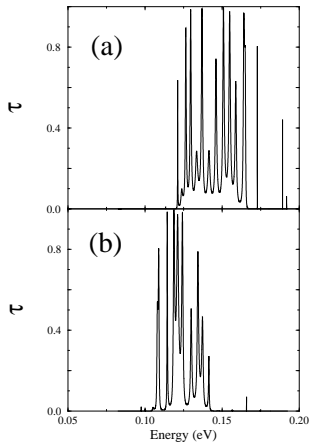}}}
\vspace*{0.2cm}
\caption{ Transmission coefficient $\tau$ versus energy E for a DSL at
(a) F = 0 and (b) $10\,$kV/cm.  The
GaAs-Ga$_{0.65}$Al$_{0.35}$As SL consists of $N=200$ barriers of $ b =
32$ \AA, whereas the thicknesses of QW are $a=32$ \AA $\,$ and $a^\prime
= 26$ \AA.}
\label{fig2}
\end{figure}

We now consider a single DQW as shown in Fig.\ \ref{fig1} in an
otherwise perfect and periodic SL. We showed analytically in Ref.\
\onlinecite{ieee} that there is an specific energy value ($E_r$) for
which the so built SL is perfectly transparent, i.e. $\tau(E_r)=1$,
where $\tau$ is the transmission coefficient.  The value of $E_r$
depends only on geometrical parameters (layer thicknesses) and it can be
fixed at the fabrication stage.  This result concerning resonant tunneling
through a single DQW in an otherwise periodic SL does not imply that
such a resonant phenomenon will survive in a disordered SL, that is,
when more than one DQW's are randomly placed in th SL. The
transfer-matrix formalism allows us to compute exactly, although not in
a closed analytical fashion, the transmission coefficient in an
arbitrary SL. An example of the behavior of the transmission coefficient
$\tau$ around the resonant energy $E_r=0.155\ldots$ is shown in Fig.\
\ref{fig2}(a) for a GaAs-Ga$_{0.65}$Al$_{0.35}$As with N=200 barriers.

We next elucidated whether the physical mechanisms giving rise to
delocalization in unperturbed systems are of relevance in the presence
of an electric field, or the presence of the field destroyed the quantum
coherence that exists at $F=0$.  To obtain the transmission coefficient
in the presence of an electric field, we develop a similar approach to
that given in Ref.\ \onlinecite{Knapp}.  As usual in scattering
problems, we assume an electron incident from the left and define the
reflection, $r$, and transmission, $t$, amplitudes by the relationships
\begin{equation}
\psi(x)=\left\{ \begin{array}{ll} e^{ik_0x}+re^{-ik_0x}, & x<0, \\
te^{ik_Lx}, & x>L,  \end{array} \right.
\label{solution}
\end{equation}
where $k_0^2=2m^*E/\hbar^2$, $k_L^2=(2m^*(E+eFL)/\hbar^2$, and $L$ is the
length of the SL. The transmission coefficient is computed as $\tau =
(k_L/k_0)|t|^2$.  Now we define $\psi(x)=(|\,t\,|\,\sqrt{k_L}) q(x)
\exp[ i\zeta(x)]$, where $q(x)$ and $\zeta(x)$ are real functions.
Inserting this factorization in Eq.\ (\ref{Schr}) we have $\zeta_x (x)
= q^{-2}(x)$ and
\begin{equation}
\left[-\,{\hbar^2\over 2m^*}\, \left( {d^2\phantom{x} \over dx^2} -
{1\over q^4(x)} \right)  +
V_{\text{SL}}(x) - eF\,x - E \right]\,q(x) = 0. \label{q}
\end{equation}
This nonlinear differential equation must be supplemented by appropriate
boundary conditions.  However, using Eq.\ (\ref{solution}) this problem
can be converted into a initial conditions equation.  In fact, it is
straightforward to prove that
\begin{equation}
\label{ic}
q(L)=k_L^{-1/2},\>q_x(L)=0,
\end{equation}
and that the transmission coefficient is given by
\begin{equation}
\tau=\,{4k_0q^2(0)\over 1+2k_0q^2(0)+k_0^2
q^4(0)+q^2(0)q_x^2(0)}.
\label{tau}
\end{equation}
Hence, we can integrate numerically (\ref{q}) with initial conditions
(\ref{ic}) backwards, from $x=L$ up to $x=0$, to obtain $q(0)$ and
$q_x(0)$, thus computing the transmission coefficient for given incoming
energy $E$ and applied voltage $V=FL$.  Figure\ \ref{fig2}(b) shows the
transmission coefficient as a function of the incoming energy for a
moderate value of the applied voltage $F=10\,$kV/cm.
We can see how the field shifts the mini-band to lower energies and
destroy some of the quasi-bound states, but an important number of them
survive.  Then we have achieved the first goal of this paper: to show
that the extended states that appear in DSL's survive in the presence of
an electric field.  Remembering that we proved previously\cite{ieee}
that this states also survive when interface roughness is taken into
account, we can conclude that the delocalization due to structural
correlations in the disorder is a very robust phenomena.  In the next
section we tackle the principal objective of this paper, namely to
present a complete dynamical study of the exciting properties of
electrons in disordered DSL.

\section{Dynamical results}
\label{sec4}

\subsection{Numerical Method}

As we mentioned in the introduction, we are interested in quantum
diffusion of wave packets under an applied electric field in
semiconductor SL's.  The equation which rules the evolution of the wave
packet is the time-dependent Schr\"odinger equation,
\begin{equation}
\label{schro}
i\hbar\frac{\partial \Psi(x,t)}{\partial t} = {\cal H}(x)\Psi(x,t),
\end{equation}
where ${\cal H}(x)$ is the single-electron Hamiltonian given in
(\ref{Schr}).  This equation has an elegant formal solution, given by
\begin{equation}
\label{elega}
\Psi(x,t) = e^{-\frac{i}{\hbar} {\cal H}(x) t} \Psi(x,0) .
\end{equation}
Using Cayley's form for the finite difference representation
of the exponential\cite{recipes}
$$
e^{-\frac{i}{\hbar} {\cal H}(x)\delta t} \simeq \frac{1- \frac{i}{2\hbar}
\,{\cal H}(x) \delta t}{1 + \frac{i}{2\hbar}  \, {\cal H}(x) \delta t},
$$
we obtain the finite-difference equation
\begin{equation}
\label{diffshro}
\left( 1 + \frac{i}{2\hbar} {\cal H} \delta t \right) \Psi_j^{k+1} =
\left( 1 - \frac{i}{2\hbar} {\cal H} \delta t \right) \Psi_j^{k},
\end{equation}
where we have replaced the wave function by its finite-difference
approximation, in time (index $k=0,1,\dots$, with $ t_k= k \delta t$)
and in space (index $j=0,1,\dots,\cal{N})$ with $x_j = j \delta x$ and
$\cal N$ the number of grid points).  We will use a centered
finite-difference approximation in $x$ for ${\cal H}(x)$ and hence we
have just a complex tridiagonal system.  This method is commonly used in
the solution of the time-dependent Schr\"odinger equation\cite{ann}
because it ensures strict norm conservation of the wave function at all
times, and the error is only of the order ($\delta t^3$).  Norm
conservation has been used at every time step as a first test of the
accuracy of results.  We use a uniformly spaced set of spatial mesh
points much larger than the SL's under consideration, and we transform
the continuous boundary conditions, which read $\Psi(\infty,t) =
\Psi(-\infty,t) = 0$, to the corresponding discrete ones
$\Psi_{-1}^k=\Psi_{ {\cal N} + 1}^k=0$.  Of course this approximation is
valid only if we choose $\cal N$ sufficiently large to make sure than
the wave function never comes close to the boundaries.  We finally note
that our initial wave function will be a superposition of plane waves of
the form
\begin{equation}
\Psi(x,0) = \left[2 \pi (\Delta x)^2\right]^{-\frac{1}{4}} \;
\exp{\left[\frac{i k_0 x - (x - x_0)^2}{4 (\Delta x)^2}\right]},
\end{equation}
where the average kinetic energy is $E=\hbar^2k_0^2/2m^*$.

\subsection{Tunneling times and other dynamical tools}

The subject of tunneling times is rich in contradictory definitions and
results.\cite{buti1,Iannacone,Iannacone2} When we measure the {\em
transmission time} $t_T$, we are trying to measure the time that a {\em
transmitted} particle spent in the SL. The transmission time is
straightforwardly obtained in the WKB limit for a ballistic electron,
\begin{eqnarray}
\label{WKB}
t_T^{\text{WKB}}(E) = \int_0^L \sqrt{\frac{m^*}{2\left(\Delta E_c - E \right)}}
 \; \chi_w(x) \; & & dx  \nonumber  \\ +
\int_0^L \sqrt{\frac{m^*}{2 E}} \; & &  \chi_b(x) \; dx,
\end{eqnarray}
where $\chi_b(z)$ and $\chi_w(z)$ are the characteristic functions of
the barriers and the wells, respectively. The mean dwell time $t_{dw}$ is
\begin{equation}
\label{tdw}
t_{dw}(E) = \int_0^{\infty} \, dt \int_0^{L}\> |\psi(x,t)|^2\,dx,
\end{equation}
and measures the average time spent by a wave packet in a given region
of space.  This time does not distinguish between particles transmitted
or reflected, and hence the mean dwell time becomes the transmission
time of a transmitted particle when most of the wave packet is
transmitted, as was pointed out by B\"uttiker and Landauer. \cite{buti1}

Numerically, it is simple to measure $t_{dw}$, and physically is a
powerful tool to measure the density of states, as can be shown
that \cite{Iannacone}
\begin{equation}
\label{rho}
\rho(E) = \frac{1}{\pi\hbar}\,t_{dw}(E).
\end{equation}

According to Ref.\ \onlinecite{Iannacone}, this relationship is only
valid for symmetrical one-dimensional structures. For non-symmetrical
structures it should be replaced by $\rho(E)=\frac{1}{2\pi\hbar}
\left[t_{dw}^r(E)+t_{dw}^l(E)\right]$, where the superscript
refers to electrons coming from the right (r) or from the left (l).
However we have found no differences between $t_{dw}^r(E)$ and $t_{dw}^l(E)$
with the parameters we are using.

Nevertheless, as Eq.(\ref{WKB}) is only valid in a perfect ballistic
regime and the mean dwell time is only the transmission time in a
idealized limit, we need to develop a method to measure $t_T$.  This
method is based on the probability $P_{T}$ that at time $t$ the particle
is found to have crossed the SL,
\begin{equation}
\label{PT}
P_{T}(t) = \int_L^{\infty} |\psi(x,t)|^2\,dx,
\end{equation}
or the probability $P_{R}$ that the particle is found to have been
reflected back by the SL
\begin{equation}
\label{PR}
P_{R}(t) = \int_{-\infty}^0 |\psi(x,t)|^2\,dx,
\end{equation}
and will be explained in the next section.

To get an estimation of the spreading of the wave packet as a function
of time we will use the time-dependent inverse participation ratio (IPR)
and the mean-square displacement ($\sigma$), defined respectively as,
\begin{mathletters}
\label{sigma}
\begin{eqnarray}
\mbox{IPR}(t) &=& \int_{-\infty}^{\infty} \, |\psi(x,t)|^4\,dx,  \\
\sigma(t) &=& \int_{-\infty}^{\infty}\> \left( x-\overline{x}\right)^2\,
|\psi(x,t)|^2\,dx.
\end{eqnarray}
\end{mathletters}
with
\begin{equation}
\overline{x} = \int_{-\infty}^{\infty} \> x \,|\psi(x,t)|^2\,dx.
\end{equation}
Usually the IPR is a good estimation of the spatial extent of electronic
states.  Delocalized states are expected to present small IPR (for long
times IPR $\sim 1/L$), while localized states have larger IPR. The
mean-square displacement is frequently also used to describe wave packet
dynamics. In the asymptotic regime ($t\to\infty$) one expects a behavior
of the form $\sigma(t) \sim t^{\gamma}$. Here $0< \gamma <1$ for
localization, $\gamma = 1$ for ordinary diffusion, $1< \gamma <2$ for
super-diffusion, and $\gamma = 2$ for ballistic regime. The later is
found in homogeneous systems\cite{Evangelou}.

\subsection{Quasi-ballistic scattering}

In this section we study the interaction of a Gaussian wave packet with
average kinetic energy $E$, with the two different classes of disordered
SL's, RSL and DSL, which we introduced in Sec.\ \ref{sec2}.  For a RSL,
we of course expect that the wave packet will be essentially reflected
for any selected energy.  However, in the case of a DSL we have two
possible scenarios.  On the one hand, if the dwell time is sufficiently
large to allow a quasi-bound state of characteristic width $\Gamma$
to be established, namely $t_{dw} \simeq \hbar/\Gamma$ (see for example
Ref.\ \onlinecite{Gennady}), we expect that
particles with energy close to the resonant one will be transmitted.
If, on the contrary, the dwell time is not sufficiently large we never
have a quasi-bound state and the behavior of the DSL will be the same
that a RSL. {\em A priori}, we have no means to decide between these two
possibilities, hence the necessity of the dynamical study we are
summarizing here to clarify whether extended states do play a role in
transport properties of DSL or not.

\begin{figure}
\setlength{\epsfxsize}{5.0cm}
\centerline{\mbox{\epsffile{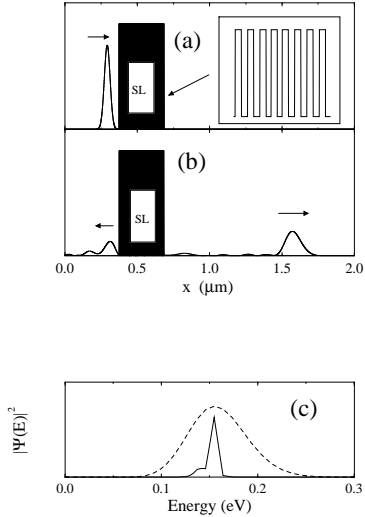}}}
\vspace*{0.2cm}
\caption{The initial probability density $| \psi(z,0) |^2$,
corresponding to a Gaussian function with $\Delta x = 200\,$\AA\ and
energy $ E=E_r=0.155\,$eV, is shown in (a).  The
potential energy $V_{\text{SL}}(x)$ is plotted as a function of $x$, for
a DSL of $N=50$ barriers of $ b =
32$ \AA, whereas the thicknesses of QW are $a=32$ \AA $\,$ and $a^\prime
= 26$ \AA. The inset in (a)
is an enlarged view of a portion of the SL potential.  The probability
density $|\psi(z,t)|^2$ at time $t = 2\,$ps is shown in (b).  The
Fourier transform for the initial wave packet (see (a)), $|
\psi(E,0)|^2$, (dashed line) and for the transmitted packet through the
DSL (see (b)) at time $t = 2\,$ps, $|\psi(E,t)|^2$, (solid line) are
shown in (c) as a function of the energy.}
\label{fig3}
\end{figure}

Figure \ref{fig3} collects the results of a typical simulation of a wave
packet for a DSL. In Fig.\ \ref{fig3}(a) we have a wave packet with a
central energy of $E=0.155\,$eV, very close to the resonant one obtained
in Sec.\ref{sec2}, traveling to impinge on a DSL. Some time afterwards,
we can see in Fig.\ \ref{fig3}(b) that a small packet has emerged in the
right part of the SL. We realize that the structure has {\em filtered}
the initial wave packet, allowing only to pass the energies laying in
the subband of extended states.  We can confirm this interpretation by
performing the Fourier transform of the emergent wave packet and
comparing it with the initial one as shown in Fig.\ \ref{fig3}(c).  We
can see the emergent wave packet has an energy spectrum much narrower
than the initial one, peaked around the resonance; this effect turns out
to be much more dramatic the larger the SL is, but we preferred to keep
within the limits of available superlattices (note that $N=50$ in Fig.\
\ref{fig3}) instead of increasing the number of wells to get more
spectacular results.

\begin{figure}
\setlength{\epsfxsize}{6.0cm}
\centerline{\mbox{\epsffile{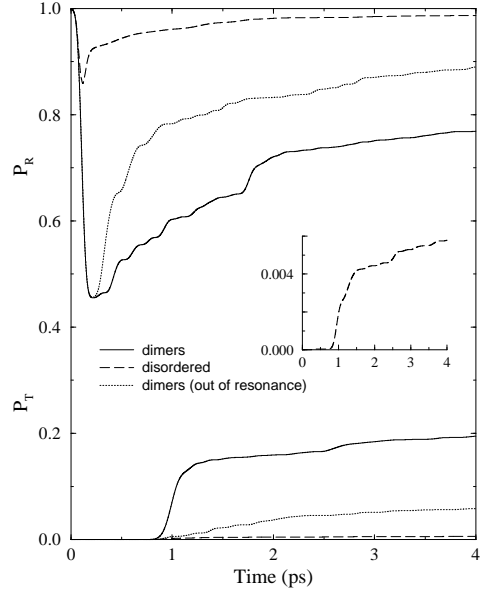}}}
\vspace*{1cm}
\caption{Dynamical transmission $P_T$ (lower curves) and reflection
$P_R$ (upper curves) probabilities are shown as a function of time for a
DSL with the same parameter as in Fig. 3, at energy $E=E_r=0.155\,$eV
(solid line) and $0.180\,$eV (dotted line) and for a RSL (with the same
parameters that de DSL) at energy $E=0.155\,$eV (dashed line).  The inset is an
enlargement of the transmission probability for a RSL, the lowest curve
in the plot.}
\label{fig4}
\end{figure}

We can understand better what is happening by looking at the dynamical
evolution of the probability of transmission (reflection) $P_{T}(E,t)$
[$P_{R}(E,t)$], i.\ e., the probability of finding the particle in the
right (left) side of the SL with energy $E$ at time $t$.  We notice that
the stationary transmission probability $\tau(E)$ which we commented
upon in Sec.\ \ref{sec2} is just the limit of $P_{T}(E,t)$,
\begin{equation}
\label{newtau}
\tau(E) = \, C \, \lim_{t \rightarrow \infty} P_{T}(E,t)
\end{equation}
where $C$ is a suitable normalization constant which depends on
$\Delta x$ and tends to unity as $\Delta x \rightarrow \infty$.
In Figure \ref{fig4},
we plot $P_{T}(E,t)$ and $P_{R}(E,t)$ as a function of time for the
resonant energy $E=0.155\,$eV (solid line) and for $E=0.180\,$eV (dotted
line) for a DSL, and for $E=0.155\,$eV (dashed line) for a RSL. For a
RSL the results are similar for any energy; we have just selected $E_r$
as a typical behavior.  For the DSL there is a great dependence of the
energy.  When we select an energy far from the resonant one we have a
behavior similar to that the RSL. However, when we choose the resonant
one, in a short interval of time $P_{T}$ reaches practically its maximum
value.  In Fig.\ \ref{fig3}(b) we can see that this fast enhancement is
due to the arrival of a compact packet corresponding to the components
of the initial wave with energies closer to $E_r$.  Again, we have
selected a small number of wells to allow an experimental verification
of this results, and we have checked that the larger the number of wells
is, the larger the differences between RSL and DSL.

\begin{figure}
\setlength{\epsfxsize}{6.5cm}
\centerline{\mbox{\epsffile{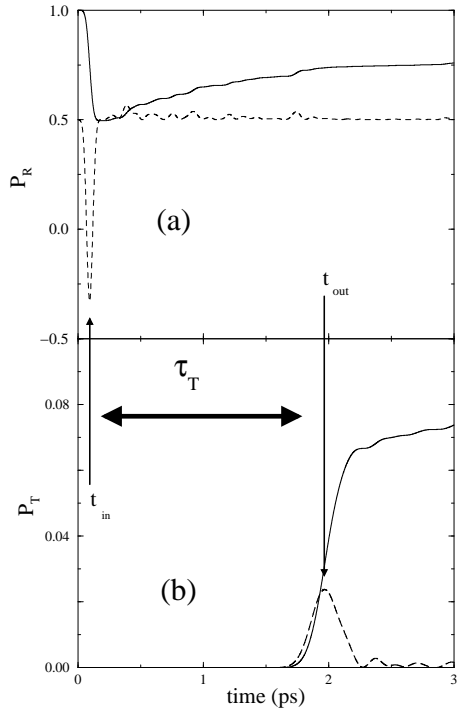}}}
\vspace*{0.5cm}
\caption{We show the typical behavior of the (a) reflection $P_R$ and
(b) transmission $P_T$ probabilities versus time for the resonant energy
in a DSL with the same parameters as in Fig.\ 3.  The dashed line
represents the derivatives (a) $\partial P_R(t,E_r) /\partial t$ and (b)
$\partial P_R(t,E_r) /\partial t$.  The arrows mark the maximum values
of the derivatives of $P_R$ and $P_T$, which we take respectively as the
initial time $t_{in}$ (i.e. when the wave packet goes inside the SL) and
the outside time $t_{out}$ (i.e. when the the wave packet goes out of
the SL). The transmission time is defined simply as $t_T =
t_{out}-t_{in}$.}
\label{fig5}
\end{figure}

We are now going to use $P_T$ and $P_R$ to find approximately the
transmission time ($t_T$).  We will choose the following convention: We
obtain the time when the wave packet entries in the SL by finding the
maximum value of $\partial P_R(t,E) /\partial t$, which indicates the
time when the probability of finding the particle inside the SL is
maximum, and we will fix our time origin $t_{in}$ at that instant (see
Fig.\ \ref{fig5}). This will be our time origin $t_{in}$.  As for the
time when the wave packet exits the SL, we can think of it as the time
when most of the particles transmitted are on the right part of the SL,
i.\ e., the time $t_{out}$ when $P_T$ is arriving to the final plateau
(cf.\ Fig.\ \ref{fig5}).  We obtain $t_{out}$ by finding the maximum
value of $\partial P_T(t,E) /\partial t$.

\begin{figure}
\setlength{\epsfxsize}{5.0cm}
\centerline{\mbox{\epsffile{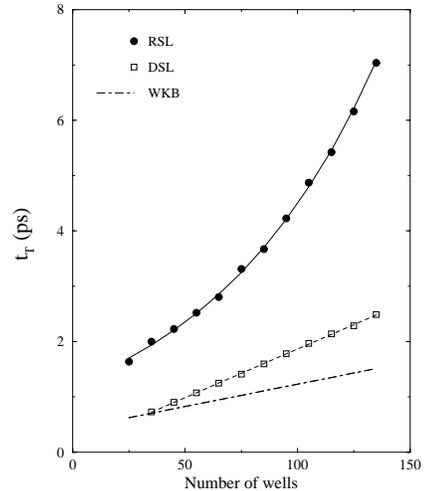}}}
\vspace*{1cm}
\caption{The transmission time $t_T$ at the resonant energy as a
function of the number of wells $N$ for a DSL (squares) with a linear
fit (dashed line) and a RSL (filled circles) with an exponential fit (solid
line).  Also depicted is the WKB approximation for the DSL case
(dot-dashed line), showing
a good qualitative agreement (same linear behavior) with the numerical
results for the DSL. Parameters for both SL's are the same as in Fig.\
3.}
\label{fig6}
\end{figure}

If the particle transmitted through the DSL is tunneling through a
ballistic channel induced by spatial correlations in the disorder, the
packet will pass the same amount of time in each well.  Therefore, the
time spent by the packet in passing through the whole SL would scale
linearly with the number of wells, i.\ e., the length of the SL. One of
the goals of this paper is to show not only that there is a significant
enhancement of the transmission probability for particular values of the
energy in DSL, but also that we are in the presence of a ballistic
transmission phenomenon in a disordered system, very close to the ideal
WKB case for periodic SLs. This conclusion can be drawn from Fig.\
\ref{fig6}.  There we have plotted the transmission time as a function
of the number of wells, for both types of SL's, namely RSL and DSL,
selecting an energy laying in the DSL mini-band.  For comparison, we
also show the results predicted by the WKB expression Eq.(\ref{WKB}).
Remarkably enough, the DSL the behavior is purely linear and very close
to that predicted by the WKB expression.  On the contrary, for the RSL
we have an exponential behavior characteristic of Anderson localized
states.  It thus becomes well established, from the dynamical view
point, the nature of the DSL as a disordered system with good transport
properties.

We now turn our attention to the mean dwell time.
When we have a particle situated in an eigenstate,
the mean dwell time in the DSL is exactly the
transmission time and has a clear relationship with the density of
states.\cite{Iannacone} On the other hand, if we have a Gaussian wave packet
the relation with the density of
states is not at all evident.  However, if we consider wave packets
with small spread in energies we can expect that Eq.~\ref{rho} continue
to hold. In Fig.\ \ref{fig7} we have plotted the
$t_{dw}$ of an initial Gaussian wave packet of energy $E=E_r$ as a
function of the number of wells.  We can see that for the DSL the
behavior is close to the linearity exhibited by the transmission time in
Fig.\ \ref{fig6} (solid line represents a linear fit).  On the contrary,
for the RSL the $t_{dw}$ exhibits a plateau because for this kind of
SL's the dwell time is dominated by $t_R$.  In Fig.\ \ref{fig2} we saw
that for a RSL most part of the wave packet just penetrates in the SL a
small number of wells; therefore $t_{dw}$ does not depend on the SL's
size as soon as the SL is larger than those few wells.  This result
agrees with the typical consistency check for any definition of
tunneling time,\cite{Iannacone2} where $t_{dw}$ is related to $t_T$ and
$t_R$ by the expression,
\begin{equation}
t_{dw} = (1-\tau) \, t_R + \tau \, t_T.
\end{equation}
For a RSL when the number of wells grows, $\tau$ goes to zero. In this
case $t_{dw}$ is equal to $t_R$, and hence the plateau observed in
Fig.\ \ref{fig7}.

{}From the preceding considerations and results, we can be confident
that in the DSL, when the transmission and dwell times behave similarly
and the propagation is quasi-ballistic, the aforementioned relationship
between peaks in the density of states corresponding to peaks in the
dwell time still holds, the only discrepancy being just a normalization
constant, related to the amount of reflected final states.  Following
this idea, we have plotted in Fig.\ \ref{fig8} the density of states,
obtained by using Eq.\ (\ref{rho}).  We can see how we have a peak in
$t_{dw}$ (and correspondingly in the density of states) for the resonant
energy, as was expected for a packet that transmits through the whole SL
by using the ballistic channel originated by the spatial correlation in
a otherwise disordered SL. We note that this coincides with the
predictions from the stationary analysis in Ref.\ \onlinecite{quasi},
thereby confirming again the results in that paper.  In particular, the
shape of the density of states curve in Fig.\ \ref{fig8} is very similar
to that obtained in our previous works.

\begin{figure}
\caption{Mean dwell time $t_{dw}$ at the resonant energy as a function
of the number of wells $N$, for a DSL (circles) and a RSL (squares).  A
linear fit for the DSL (solid line) is also shown.  Parameters for both
SL's are the same as in Fig. 3.}
\label{fig7}
\end{figure}

To conclude this zero field study of the dynamics of DSL's, we close the
section by studying the spreading of the initial wave packet versus time
for both kinds of SL's.  In Fig.\ \ref{fig9} we have plotted on a
log-log scale the mean-square displacement $\sigma$ as a function of
time for a wave packet of energy $E=E_r$ incident in a DSL (solid line)
and in a RSL (dot-dashed line), and for energy $E=0.180\,$eV impinging
on a DSL (dashed line).  At short times we can see a practically
constant behavior of $\sigma$ which can be associated to the period
while the wave packet is traveling towards the SL's.  When the packet
hits the SL, we see during a short time a decrease of $\sigma$, followed
immediately by a rapid increasing of this magnitude.  The decreasing is
a consequence of the tails of the wave packet reaching the leading part,
which is being retained by its collision with the SL; once the whole
packet is interacting with the SL, the behavior is close to power-like.
We have fitted the results showed in Fig.\ \ref{fig9} for times
larger than $1$ ps to a function of the form $\sigma(t) \sim t^{\gamma}$.
We have obtained for the DSL at the resonant energy
$\gamma = 1.120 $, i.e., we are in a super-diffusive regime, whereas  for
the DSL away from
the resonant energy $\gamma = 1.000$, right at the limit
between a localized regime and a ordinary diffusive one. Finally, for
the RSL $\gamma = 0.837$, indicating that we are clearly
in a localized regime.
The $\sigma$ for the RSL is always much lower than for the DSL and
increases with a larger approximate exponent, i.\ e., after some time
($\sim 1\,$ps) the packet is much more localized for the RSL than for
the DSL (even out of the resonance).  This is evidently a consequence of
localization effects coming from the uncorrelated disorder of the RSL,
and whose influence is much less in the DSL case given the availability
of extended states.  Such a phenomenon further confirms the conclusions
we have been drawing all along this section.  We will come back to these
results in the next section, when dealing with electric field effects.

\begin{figure}
\setlength{\epsfxsize}{4.7cm}
\centerline{\mbox{\epsffile{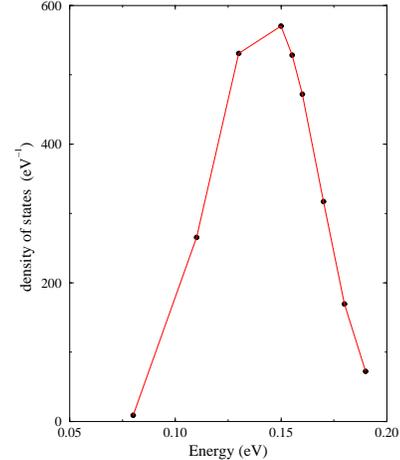}}}
\vspace*{0.2cm}
\caption{Density of states $\rho$ as a function of the incident energy
$E$ obtained measuring $t_{dw}$ of a Gaussian wave packet of energy $E$
in a DSL with the same parameters as in Fig.\ 3.  Line is only a guide
to the eye.}
\label{fig8}
\end{figure}

\begin{figure}
\setlength{\epsfxsize}{4.7cm}
\centerline{\mbox{\epsffile{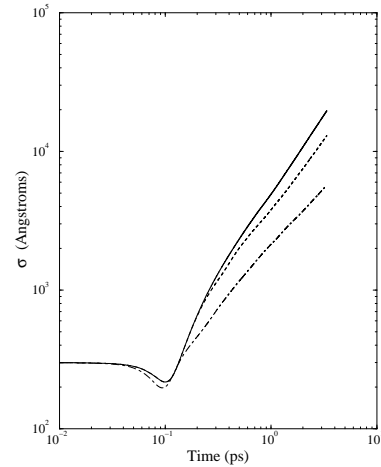}}}
\vspace*{0.2cm}
\caption{Mean-square displacement $\sigma$ of an initial Gaussian wave
packet incident in a DSL at the resonance energy $E_r=0.155\,$eV
(solid line), and at energy $E = 0.180\,$eV (dashed line) and for a
RSL at energy $E = 0.155\,$eV (dot-dashed line).  Parameters for both
SL's are the same as in Fig.\ 3.  Note the log-log scale.}
\label{fig9}
\end{figure}

\section{Electric field effects}
\label{sec5}

\begin{figure}
\setlength{\epsfxsize}{7.0cm}
\centerline{\mbox{\epsffile{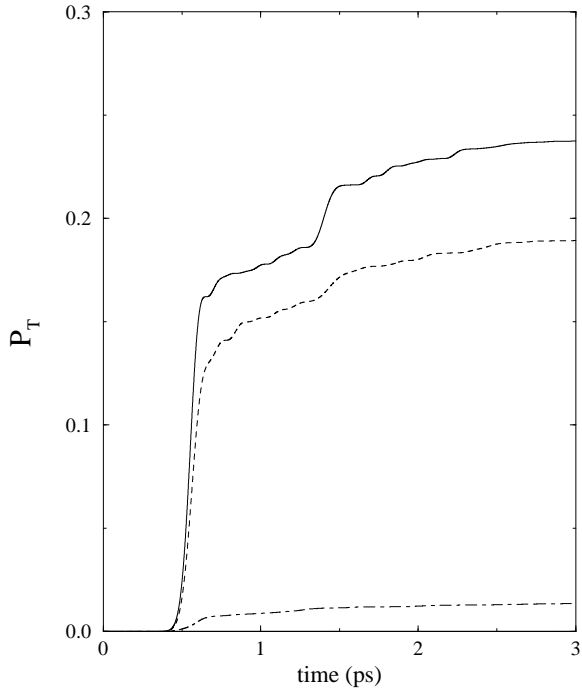}}}
\vspace*{0.2cm}
\caption{Transmission probability $P_T$ as a function of time for a DSL
with the same parameters as in Fig.\ 3, at energy $E=E_r=0.155\,$eV,
for different values of the electric field $F = 1\,$kV/cm
(dashed line) and $5\,$kV/cm (dot-dashed line).  For
comparison we also show the result for $F = 0$ (solid line).}
\label{fig10}
\end{figure}

The zero field simulations we have been presenting provide an incomplete
picture of electron dynamics in DSL's, as in this case it is obvious
that technologically applicable phenomena would involve electric fields.
Therefore, in this section we study the dependence of the dynamical
characteristics discussed in the previous paragraphs on an electric
field.  In Sec.\ \ref{sec2} we explained that the subband of extended
states appearing in DSL's is shifted to lower energies and reduces its
width in the presence of moderate electric fields.  We want to confirm
that the correspondingly shifted quasi-bound states will have time
enough to be established in the dynamical interaction of a Gaussian wave
packet with a DSL potential when there is an applied field.  To this
end, in Fig.\ \ref{fig10} we have plotted $P_T$ for a Gaussian wave
packet incident on a DSL with $E=0.155\,$eV, for different values of the
electric field $F$.

\begin{figure}
\setlength{\epsfxsize}{7.0cm}
\centerline{\mbox{\epsffile{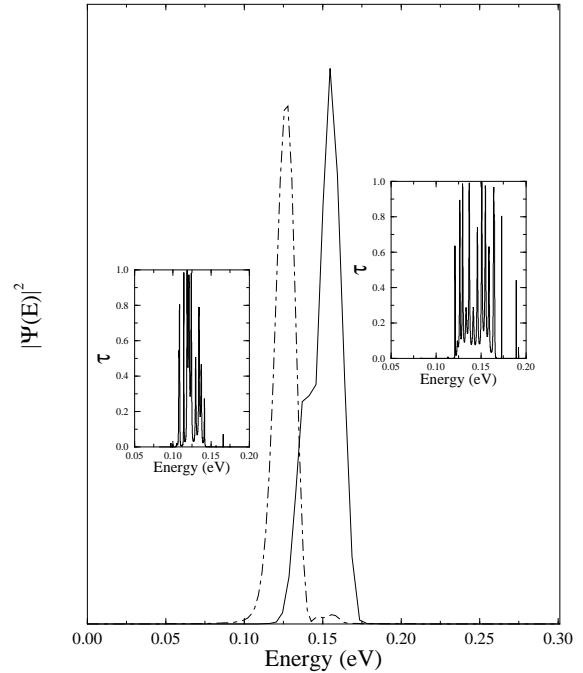}}}
\vspace*{1cm}
\caption{The Fourier transform for the transmitted packet versus energy
in the presence of an electric field $ F = 10\,$kV/cm
(dashed line); solid line shows the result at
zero field.  The insets show the transmission coefficient $\tau$ as a
function of energy for the two values of the electric field $F$ for
reference.  The SL parameters are the same as in Fig.\ 3}
\label{fig11}
\end{figure}

We can see that, at least for moderate fields, the dwell time is large
enough ($t_{dw} \simeq \hbar/\Gamma$)
to allow the quasi-bound states to be established thus permitting
the transmission of the resonant components of the packet.  The maximum
value of $P_T$ decreases with the field due to the shift of the
mini-band and because the miniband becomes much narrower the larger the
electric field applied is.  Interestingly, we can still find a {\em
dynamical} resonant energy looking at the Fourier transform of the
transmitted wave packet, as shown in Fig.\ \ref{fig11}.  We want to
stress that these results can be of interest for applications, because a
DSL turns out to be a structure that works like an adaptive electronic
filter, namely, by tuning properly the SL parameters we can filter the
energies contained in a narrow band.  Moreover, this band can be
displaced to the desired values by selecting a particular value of the
applied electric field.

\begin{figure}
\setlength{\epsfxsize}{7.5cm}
\centerline{\mbox{\epsffile{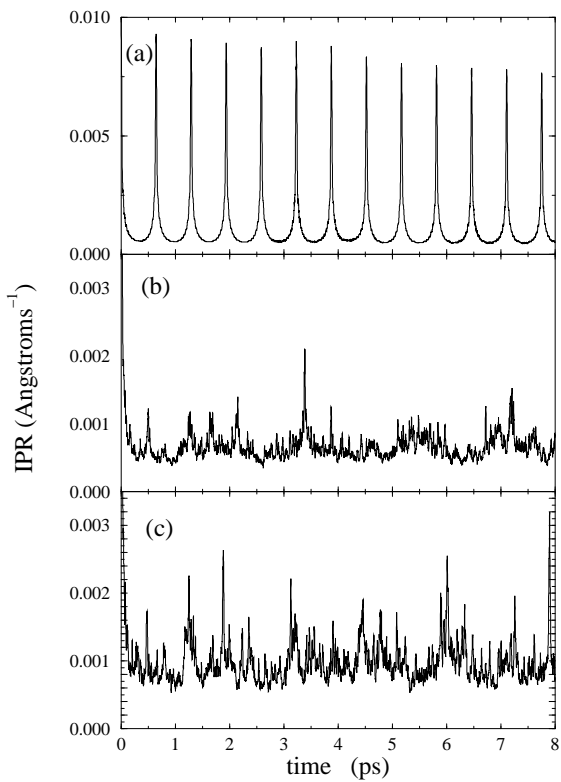}}}
\vspace*{1.5cm}
\caption{IPR of an initial Gaussian wave packet placed in a (a) perfect
SL, (b) DSL, and (c) RSL as a function of time.  For the three kind of
SL's, the number of wells is $N = 200$ and the rest of parameters are
the same as in Fig.3.}
\label{fig12}
\end{figure}

\begin{figure}
\setlength{\epsfxsize}{6.0cm}
\centerline{\mbox{\epsffile{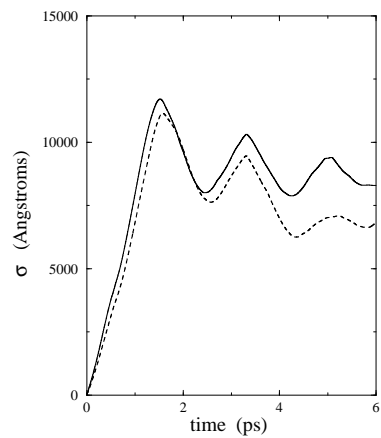}}}
\vspace*{1cm}
\caption{Mean-square displacement $\sigma$ of an initial Gaussian wave
packet placed in a DSL (solid line) and RSL (dashed line), in the
presence, of an electric field $F = 10\,$kV/cm, as a
function of time.  SL parameters are the same as in Fig. 12}
\label{fig13}
\end{figure}

There is another aspect of the influences of electric fields on electron
dynamics which is worth considering, namely the following: It is well
known that, when an electric field is applied to a periodic SL, the
localization of the initially extended states produces an oscillatory
behavior of the wave packet, the so-called Bloch oscillations.  Of
course, Bloch oscillations require a quasi-perfect quantum coherence and
a perfectly defined phase to be self-sustained in time.  This is not the
case in a DSL where electronic states increment their phase by a factor
$\pi$ whenever they pass over a DQW. \cite{anxo1} The question then
arises as to what will the corresponding phenomenology in this case be.
In order to answer this question, in Fig.\ \ref{fig12} we plot the IPR,
defined in Eq.\ (\ref{sigma}), as a function of time for (a) a perfect
SL, (b) a DSL and (c) a RSL, in the presence of an electric field.  The
initial condition of these simulations was that, at $t=0$, we placed a
Gaussian wave packet with an energy of $E=0.155\,$eV and $\Delta x =
20 \,$\AA\ in the center of each one of those SL's with $50$ wells.  In
our case $F=10\,$kV/cm, $d = a + b = 64\,$\AA.  In an
ordered SL, the Bloch period will then be $T_{\text{Bloch}} = h/eFd
\sim 0.646\,$ps, in perfect agreement with obtained from Fig.\
\ref{fig12}(a).  For the disordered superlattices there is an
oscillatory behavior at the beginning but in a short time the IPR
achieves a randomly fluctuating, but stationary in mean, value.  This
indicates the existence of decoherence effects in both disordered
lattices, the difference between the RSL and the DSL being that the
latter shows a smaller mean value of the IPR, in agreement with the less
localized character of its states.  The remnants of oscillatory behavior
for the disordered SL's are more clearly characterized by looking at the
mean square displacement $\sigma$ as a function of the time, which is
shown in Fig.\ \ref{fig13}.  Again, we can see how in the DSL the wave
packet is much more delocalized that in the RSL as a consequence of the
presence of a narrow band of extended states.

\section{Discussion and Conclusions}
\label{sec6}

In this work we have successfully shown that the good transport
properties predicted by previous static studies of SL's with correlated
disorder give rise to corresponding dynamical phenomena of interest.  To
this end, we have reported on dynamical properties of electrons in
intentionally disordered SL's, computed by using high-accuracy numerical
methods to solve the time dependent Schr\"odinger equation for the
complete Hamiltonian.  In this respect, as the two main global
conclusions of the present paper, we want to stress, first, that the
dynamical results we show prove independently the existence of extended
states with physical consequences in disordered systems, and second,
that the validity of our previous static calculations in Refs.\
\onlinecite{ieee} and \onlinecite{quasi} to characterize electron
transmission through nanostructures has been set on firm grounds due to
its perfect agreement with the dynamical analysis.

Aside from the above general conclusion, which we draw from the
consideration of a number of dynamical tools, we would like to summarize
a few aspects more that we have learned from our simulation program.  In
particular, we have proposed a method to find the transmission time by
using the temporal transmission probability, provided that such probability
presents abrupt changes as a function of time. By means of this procedure
to compute the transmission time, we have been able to show that the
propagation of electrons with energies in the subband of extended states
of a DSL is ballistic, very similar to that of ordered SL's.  This is a
dramatic manifestation of delocalization by correlations, more so when
compared to the exponential growth of the transmission time we have
obtained for usual RSL's.  In that regime, we have shown that the
relationship between dwell times and density of states,
holds for Gaussian wave packets, by computing the density of states and
finding the same result as in our stationary calculations.
Interestingly, the fact that correlations do not impede the phase
decoherence of the wavepacket, the properties depending on symmetries of
the system (translational invariance) are not recovered.  This is the
case, e.\ g., of Bloch oscillations.  In any event, measurements of the
IPR point out once more the differences between DSL's and RSL's.  All
this characterization is confirmed by measuring the mean square
displacement of electrons, which are seen to evolve faster in DSL's.
Finally, we have also confirmed that low to moderate electric fields do
not destroy the transport properties of DSL's, which is very important
if DSL's are to be built and used for any practical purpose.

To conclude, a few words are in order regarding possible applications of
the present work.  It seems quite clear to us that several of our
results can be useful for nanotechnological devices with specific,
special features.  To begin with, the great difference of transmission
times between extended states and localized ones may provide a powerful
tool for measure the extended character of the states in open systems.
Besides, it can also be used to measure the amount and character of the
disorder inherently present in any {\em periodic} SL, by obtaining the
width of the band of extended states in the actual SL and comparing it
to the theoretically predicted one.  However, what we think by far is
the most promising application of DSL in nanotechnology has to do with
their filter-like behavior.  We have seen that it is possible, by means
of an applied electric field, to control the center and width of the
band of extended states, therefore allowing for a tunable filtering of
wavepackets, i.\ e., of electrons.  This capability, present already in
practically achievable DSL's of some 50 wells, can be used to design a
new family of electronic devices.  In this respect, it is quite clear
that a natural extension of this work would be to study the interaction
of RSL and correlated disordered SL's, with an AC-electric field, using
the complete Hamiltonian.  Preliminary tight-binding results
\cite{Holthaus} appears to show exciting new phenomena in these
structures.  We envisage that appropriate choices of the frequency
and/or intensity of the field can give rise to crucial changes in the
filtering properties of DSL's.  Further work along these lines is
currently in progress. \cite{moco}

\acknowledgments

Work at Legan\'es and Madrid is supported by the Comisi\'on
Interministerial de Ciencia y Tecnolog\'\i a (CICyT, Spain) under Grant
No.\ MAT95-0325.  G.\ P.\ B.\ gratefully acknowledges partial support
from Linkage Grant No.\ 93-1602 from NATO Special Programme Panel on
Nanotechnology.

\end{multicols}

\end{document}